\documentclass{aims}
\usepackage{amsmath}
  \usepackage{paralist,tikz}
\usepackage[T1]{fontenc}
  \usepackage{graphics} 
  \usepackage{epsfig} 
\usepackage{graphicx}  \usepackage{epstopdf} 
 \usepackage[colorlinks=true]{hyperref}
\hypersetup{urlcolor=blue, citecolor=red}

\newtheorem{proposition}{Proposition}

\theoremstyle{definition}

\newtheorem{remark}{Remark}

\DeclareMathOperator*{\dt}{\delta t}
\DeclareMathOperator*{\R}{\mathbb{R}}

\makeatletter
\renewcommand{\fnum@figure}{\small \textbf{\figurename~\thefigure}}
\renewcommand{\fnum@table}{\small \textbf{\tablename~\thetable}}

\newcommand{\ba}{\begin{array}{ll}}
\newcommand{\ea}{\end{array}}

\newcommand{\be}{\begin{equation}}
\newcommand{\ee}{\end{equation}}

\newcommand{\bi}{\begin{itemize}}
\newcommand{\ei}{\end{itemize}}

\newcommand{\bc}{\begin{center}}
\newcommand{\ec}{\end{center}}

\newcommand{\bfig}{\begin{figure}[!ht]}
\newcommand{\efig}{\end{figure}}

\newcommand{\ben}{\begin{enumerate}}
\newcommand{\een}{\end{enumerate}}

\newcommand{\bmat}{\left[\begin{matrix}}
\newcommand{\emat}{\end{matrix}\right]}

\title[Modelling pedestrian dynamics with port-Hamiltonian systems] 
      {Multi-scale description of pedestrian collective dynamics with port-Hamiltonian systems$^*$} 

\author[Antoine Tordeux and Claudia Totzeck]{}

\subjclass{Primary: 76A30, 34C60; Secondary: 82C22.}
 \keywords{Pedestrian dynamics, force-based model, port-Hamiltonian system, collective dynamics, Hamiltonian order parameter, multiscale description.}

 \email{tordeux@uni-wuppertal.de}
 \email{totzeck@uni-wuppertal.de}

\thanks{The first author is supported by NSF grant xx-xxxx}

\thanks{$^*$Corresponding author: First-name1 last-name1}

\begin{document}
\maketitle

\centerline{\scshape Antoine Tordeux}
\medskip
{\footnotesize
 \centerline{School of Mechanical Engineering and Safety Engineering}
   \centerline{University of Wuppertal}
   \centerline{Wuppertal, Germany}
} 

\medskip

\centerline{\scshape Claudia Totzeck}
\medskip
{\footnotesize
 \centerline{School of Mathematics and Natural Sciences}
   \centerline{University of Wuppertal}
   \centerline{Wuppertal, Germany}
}

\bigskip

 \centerline{(Communicated by the associate editor name)}

\begin{abstract}
Port-Hamiltonian systems (PHS) theory is a recent but already well-established modelling approach for non-linear physical systems. Some studies have shown lately that PHS frameworks are relevant for modelling and control of swarm and multi-agent systems. 
We identify in this contribution a general class of microscopic force-based pedestrian models that can be formulated as a port-Hamiltonian system. 
The pedestrian PHS has linear structure and dissipation components. 
Non-linear effects come from isotropic pedestrian interactions.
Simulation results on a torus with disordered initial states show that the port-Hamiltonian pedestrian model can exhibit different types of dynamics. 
They range from relaxed speed models with no interaction, dynamical billiards, or crystallization dynamics to realistic pedestrian collective behaviors, including lane and strip formation for counter and crossing flow. 
The port-Hamiltonian framework is a natural multiscale description of pedestrian dynamics as the Hamiltonian turns out to be a generic order parameter that allows us to identify specific behaviours of the dynamics from a macroscopic viewpoint.  
Particular cases even enable through energy balance to determine the Hamiltonian behavior without requiring the tedious computation of the microscopic dynamics. 
Using PHS theory, we systematically identify a critical  threshold value for the Hamiltonian, which relies only on exogenous input and can be physically interpreted. 
\end{abstract}

\section{Introduction}\label{introduction}

\begin{figure}[t]
\centering\bigskip\small
\input{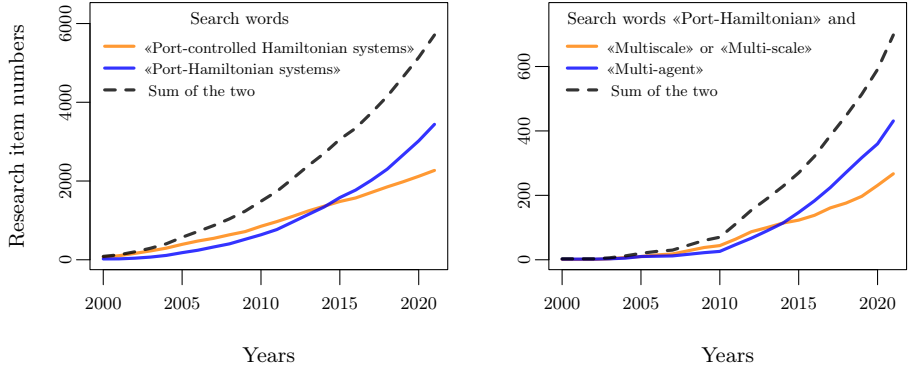}\vspace{-4mm}
\caption{Cumulative research item numbers from 2000 to 2021 for the search words \guillemotleft Port-controlled Hamiltonian systems\guillemotright\ and \guillemotleft Port-Hamiltonian systems\guillemotright\  (left plot) and the search words \guillemotleft Port-Hamiltonian\guillemotright\ and \guillemotleft Multi-agent\guillemotright\  or \guillemotleft Multiscale\guillemotright\ (right plot).
Requests done on googlescholar \protect\cite{googlescholar} the 22.08.2022.}
\label{fig:Intro}
\end{figure}

Port-Hamiltonian systems (PHS) are recent modelling approaches for nonlinear physical systems \cite{van2006port,van2014port}. 
They date back to the 1980s and the pioneer works of Arjan van der Schaft and Bernhard Maschke on Hamiltonian systems with inputs and outputs \cite{van1981symmetries,MaschkeVdsBreedveld}. 
The modelling concept, initially called \textit{Port-Controlled Hamiltonian Systems}, was established during the 1990s \cite{maschke1993port}. 
The terminology \textit{Port-Hamiltonian Systems} was democratised during the 2000s \cite{van2006port,van2014port} and is nowadays increasingly predominant (see Fig.~\ref{fig:Intro}). 
While Hamiltonian systems are often referred to as conservative systems, PHS enable for dissipation and their ports can be used for control actions and external factors influencing the dynamics.
The modelling approach also allows for straightforward computation of the system output and Hamiltonian behavior. 

To fix the setting and some standard PHS terms we recall the structure of a PHS system with Hamiltonian $H \colon \R^n \rightarrow \R$ as follows:
\begin{align*}
\dot z(t) = (J-R) \nabla H(z(t)) + Bu(t), \quad y(t) = B^\top \nabla H(z(t)),
\end{align*}
where $J\in \R^{n\times n}$ is skew-symmetric, $R\in \R^{n\times n}$ positive semi-definite, $B \in \R^{n\times m}$ with \textit{state} $z$ and \textit{output} $y$. In most of the examples the input $u(t)\in \R^m$ is lower dimensional than the state, i.e.~$m\ll n.$ For $B\equiv 0$ and $R\equiv 0$ we obtain a conservative Hamiltonian system. $R>0$ yields \textit{dissipation} and the input $u$ allows to feed energy into the dynamics. This can be understood mathematically with a simple calculation. Indeed, with the skew-symmetry of $J$ and the positive semi-definiteness of $R$, it holds
\begin{align*}
\dot H(z(t)) &= \nabla H(z(t)) \cdot \dot z(t) \\&= \nabla H(z(t)) \cdot \left( (J-R) \nabla H(z(t)) + Bu(t) \right) \\&= - \nabla H(z) \cdot R \nabla H(z(t)) + \nabla H(z(t)) \cdot B u(t) \\ &\le y(t) \cdot u(t)
\end{align*}
In the classical Hamiltonian framework the skew-symmetry yields a zero right-hand side with means the system is conservative. For $R>0$ we obtain an inequality which accounts for the dissipation. $u \ne 0$ allows to feed in or pull out energy, making the system \textit{reactive}. As the input and output allows for coupling of PH systems and to exchange information, they are denoted \textit{ports}.

PHS were initially designed for network and graph modelling \cite{van2004port,van2013port}. 
Nowadays, various physical domains, including thermodynamics, electromechanics, electromagnetics, fluid mechanics, or hydrodynamics, can be tackled using PHS (see the review \cite{rashad2020twenty}).
Indeed, the PHS functional structure, mitigating the modelling between conserved quantities, dissipation, input and output, is a meaningful representation of many systems. 
Recent results have shown the relevance of PHS for self-propelled multi-agent systems modelling and control \cite{knorn2015overview}.  
Such a recent expansion partly comes from the analogy between multi-agent systems and (dynamic) graphs \cite{knorn2016deviation}.
Examples of port-Hamiltonian multi-agent systems arise in reliability engineering, for complex mechanical systems \cite{wang2016output,cristofaro2022fault}, consensus and opinion formation \cite{van2010Consensus,xue2019opinion}, 
multi-input multi-output multi-agent systems \cite{sharf2019analysis}, swarm behaviors \cite{matei2019inferring}, 
or autonomous vehicles, e.g., path-tracking \cite{ma2021path} or modelling and safety analysis of adaptive cruise control systems \cite{knorn2014passivity,dai2017safety,dai2020safety}. 
Such microscopic agent-based modelling approaches rely on PHS by ordinary, stochastic, or delayed differential equation systems. At macroscopic scales, traffic flow models \cite{bansal2021port} and fluid dynamics models \cite{rashad2021port} 
are based on infinite-dimensional PHS by partial differential equations using Stockes-Dirac structures as bond graph representations.
In all the modelling approaches, the Hamiltonian quantifies a (generalized) total energy of the system. 

Besides structural advantages and preservation of physical quantities, technical benefits of PHS yield well-posedness and uniqueness properties of the solutions, energy balance, and a well-established stability analysis using the Hamiltonian as Lyapunov function 
\cite{van2006port,van2014port}.

Classical pedestrian models rely on three behavior levels. At the strategic level, pedestrians decide on their activity and associated travel needs. The tactical level concerns short-term decisions, e.g., choosing the route  by considering obstacles and local pedestrian densities. The operational level describes the pedestrian motion resulting from  interaction with immediate surroundings, for example, collision avoidance while maintaining a steady course.
The processes at strategic and tactical level are usually considered as exogenous control factors in operational pedestrian simulations. 

Operational pedestrian models exist since the 1970s and were introduced in the pioneering works by Hirai and Tarui on microscopic simulations \cite{hirai1975simulation} and by Henderson on macroscopic crowd representation \cite{Henderson1971a}. 
Nowadays, the modelling approaches range from macroscopic, mesoscopic, and microscopic models, among other modeling scale characteristics \cite{martinez2017modeling,Chraibi2018}. 
Macroscopic and mesoscopic approaches come from continuous fluid dynamics or gas-kinetic models describing aggregated flow behaviors in Eulerian domains, while microscopic approaches reproduce individual pedestrian motions in Lagrangian frameworks. 
Many reviews focus on pedestrian modeling scales and passages from one modeling level to another  \cite{bellomo2011modeling,albi2019vehicular,bellomo2022towards}. 
Indeed, mesoscopic or macroscopic representations are in practice easier to analyse than multi-agent microscopic systems.
Hydrodynamic and mean-field limits are some of the most used techniques \cite{albi2019vehicular,Fischer2020,Barre2020,burger2021mean,burger2021meanb}.
Numerous studies rely on individual pedestrian motions with microscopic continuous approaches, notably with the social force model (SFM) \cite{Helbing1995}.
The social force model is an inertial, second-order operational approach belonging to the force-based model class \cite{chraibi2011force}. 
Speed-based models inspired by robotics, e.g., the reciprocal velocity avoidance (RVO) and optimal, reciprocal collision avoidance (ORCA) first-order models \cite{vanderBerg2008,vanderBerg2011}, are also relevant in the field, notably for computer graphics animation \cite{van2021algorithms}.
One of the advantages of microscopic approaches compared to macroscopic ones is their natural ability to reproduce heterogeneous behaviors and notably mixed flow of pedestrians with different desired directions. 
The challenge consists of identifying microscopic underlying mechanisms and interaction types initiating the emergence of collective phenomena such as lane and strip formation for counter and crossing flow, respectively. 
The emergence of collective dynamics can be detected by stability analysis \cite{albi2019vehicular,Barre2020,chraibi2015jamming}
or with the help of order parameters \cite{nowak2012quantitative,khelfa2022heterogeneity}. 

In this contribution, we identify a generic class of force-based pedestrian models that can be formulated as a port-Hamiltonian system by a second-order ordinary differential equation system. In contrast, to the  modelling approach has been proposed recently for the Cucker-Smale swarm model \cite{matei2019inferring}, the dynamics here is not only self-propelled but allows for desired velocities or paths of the pedestrians.
PHS is found to be a pertinent modelling framework for pedestrian behaviors by decomposing the dynamics between interaction, dissipation, and control, i.e., input and output. 
A modelling parallel between force-based pedestrian models and the port-Hamiltonian formulation is discussed. 
Besides explicit control of the dynamics through the agents' input ports, the PHS formulation has the advantage of a direct multiscale framework since the aggregated information of the Hamiltonian can be used to define a macroscopic order parameter. 
In the literature, the order parameters are generally tailored to identify specific collective dynamics. 
Here, the systematic quantification of system energy through the Hamiltonian offers a generic order parameter to analyse pedestrian collective dynamics.
In particular cases, energy balance allows determining the Hamiltonian behaviour  without requiring the tedious computation of the microscopic dynamics. 
This makes port-Hamiltonian systems a promising modelling approach for pedestrian dynamics that we aim to explore further.

The manuscript is organised as follows: 
we present the microscopic force-based pedestrian model, its port-Hamiltonian formulation, and appropriate numerical discretization schemes in the next section. 
Simulation results on a torus with disordered initial states show that the port-Hamiltonian pedestrian model can describe different types of dynamics in Sec.~\ref{simulation}.
We first analyse specific cases for which repulsive interaction, dissipation, or input are zero to illustrate the role of the different model components on the dynamics. 
Collective pedestrian dynamics, including lane and strip formation for counter and crossing flow, and their characterisation using an order parameter based on the Hamiltonian are shown hereafter. 
Sec.~\ref{conclusion} provides a conclusion and further perspectives for port-Hamiltonian pedestrian modelling.

\section{Port-Hamiltonian pedestrian model}\label{definition}
We begin with the presentation of the details of the microscopic pedestrian model. Then the dynamics is rewritten to show its port-Hamiltonian structure. The section concludes with a discussion of different scenarios for the evolution of the Hamiltonian over time.
\subsection{Microscopic dynamics}

We consider $N\ge2$ particles on a torus with position 
$$q_i \colon [0,T] \rightarrow \mathbb R^2$$ 
and momentum
$$p_i \colon [0,T] \rightarrow \mathbb R^2$$ 
for $i=1,\dots,N.$ By $$Q_{ij}(t)=q_i(t)-q_j(t) \in \mathbb R^2,$$ 
we denote the relative position of pedestrian $i$ to pedestrian $j$ at time $t\in[0,T]$. For notational convenience, we collect the relative positions of pedestrian $i$ in the vector, $$Q_i=(Q_{ij})_{j\not=i} \colon [0,T] \rightarrow \mathbb R^{2\cdot (N-1)} \text{ for } i,j=1,\dots,N,\; j\ne i.$$  Throughout the article, we assume that the pedestrians have nomalized mass, i.e.\  $m=1$, hence momentum and velocity coincide.
Following the classical modelling approach of force-based models \cite{Chraibi2018,Helbing1995,chraibi2011force}, we identify two main components in operational pedestrian dynamics: 
\begin{enumerate}
    \item isotropic short-range repulsion among neighboring pedestrians;
    \item relaxation towards an exogenous desired velocity.
\end{enumerate} 
In a minimal model, the desired velocity, i.e., desired direction (vector) and desired speed (scalar), is assumed exogenous and provided at a tactical modelling level. The microscopic model dynamics for the $i$-th pedestrian reads
\begin{equation}\label{modmicro}
   \begin{cases}
         ~\dot Q_{ij}(t)=p_i(t)-p_j(t),\; j=1\dots,N, j\ne i & \qquad Q_{ij}(0)=Q_{ij}^0,\\[1mm]
         ~\dot p_i(t)= \lambda\big(u_i(t)-p_i(t)\big)-\sum_{j\not=i}\nabla U\big(Q_{ij}(t)\big), & \qquad p_i(0)=p_i^0,
    \end{cases}
\end{equation}
with
\begin{itemize}
\item exogenous desired velocities 
$$u(t)=(u_1(t),\ldots,u_N(t))\colon [0,T] \rightarrow \mathbb R^{2\cdot (N-1)}$$
coming from a tactical model. Even though the desired velocities may be time-dependent in general, they are assumed constant in the following. 
\item relaxation rate $\lambda\ge0$ modelling the sensitivity w.r.t.~the desired velocity;
\item distance-based non-linear repulsive interaction potential given by 
\begin{equation}
    U:\mathbb R ^2\mapsto\mathbb R_+, \qquad U(x)=AB e^{-|x|/B},
\end{equation} 
where $|\cdot|$ is the minimal distance on the torus. The potential $U$ is isotropic since it does not depend on the current velocity. The assumption of distance-based repulsion with the neighbors comes from proxemics social concepts and is also safety-motivated as it allows the pedestrians to avoid collisions. 

\end{itemize}
Note that 
\begin{equation}\label{eq:odd}
    \nabla U(x)=-\frac x{|x|}Ae^{-|x|/B}=-\nabla U(-x)
\end{equation} is an odd function and the  implicit assumption that interactions depend only on the distances of two interacting agents. Both will be crucial in the derivation of the PHS formulation later on. Here, we propose to model forces with the help of a gradient of a nonlinear distance-based potential as for example in \cite{Totzeck2020}. 
In general, any repulsive distance based potential with odd potential can be used.
Altogether, the model corresponds to a simplified version of the well-known \emph{social force model} (SFM) by Helbing \& Moln\`ar (1995) \cite{Helbing1995}. The main simplification is that the anisotropic vision field mechanism is neglected.
Moreover, we remark that introducing $Q_{ij}$ increases the number of variables. Nevertheless, we note that the information in the system is not altered as the information of each $Q_i$ can be computed knowing one of the other $Q_j$'s.

\subsection{Port-Hamiltonian formulation}

It has been shown recently that the Cucker-Smale swarm model can be represented as a port-Hamiltonian system \cite{matei2019inferring}. 
The formulation is conjectured for large systems and demonstrated with three interacting particles. 
It turns out that the microscopic model Eq.~(\ref{modmicro}) for $N$ agents can also be formulated as a port-Hamiltonian system. This is the main step towards the formulation of the Hamiltonian-based order parameter which is the main result of this article.

\begin{proposition}\label{prop1}
Port-Hamiltonian formulation of the microscopic pedestrian system Eq.~(\ref{modmicro}).\smallskip

\noindent Denoting $Q=(Q_{1},\ldots,Q_{N})$, 
$p=(p_1,\ldots,p_N)$, and 
$z=(Q,p)^\text T$, a port-Hamiltonian formulation of the microscopic pedestrian system Eq.~(\ref{modmicro}) is given by
\begin{equation}\label{PHS}
\begin{cases}
    ~\dot{z}(t)=(J-R)\nabla H(z(t))+\lambda \tilde u(t),\qquad z (0)=(Q^0,p^0),\\[1.5mm]
    ~y(z(t))=\lambda\nabla H(z(t)),
\end{cases}
\end{equation}
with
\begin{equation}
    J=\left[\begin{matrix}
    0& M\\
    -M^\text T& 0
    \end{matrix}\right],\quad
    R=\left[\begin{matrix}
    0&~0\\
    0&~\lambda I
    \end{matrix}\right],\quad
		M=\left[\begin{array}{c}M_1\\ \vdots\\M_N\end{array}\right],
\end{equation}
\begin{equation}
    \tilde u(t)=\left[\begin{matrix}0\ldots0&u_1(t)\ldots u_N(t)\end{matrix}\right]^\text T,
\end{equation}
and
\begin{equation}
M_1=\left[\begin{matrix}
    1&-1&0&0&\ldots&0\\
    1&0&-1&0&\ldots&0\\
		\vdots&&&&&\vdots\\
		1&&&&&-1
    \end{matrix}\right],\qquad
M_2=\left[\begin{matrix}
    -1&1&0&0&\ldots&0\\
    0&1&-1&0&\ldots&0\\
		\vdots&&&&&\vdots\\
		0&1&&&&-1
    \end{matrix}\right],\qquad ...
\end{equation}
Here, $J,R\in\mathbb R^{N(N-1)\times N}$, $J$ is skew-symmetric and admits block structure while $R$ is positive semi-definite. The input $u$ allows for control actions and $y$ is the output. 
The Hamiltonian $H$ is the sum of the distance-based repulsive potentials and quadratic speeds, i.e.~the sum of kinetic and potential energy
\begin{equation}\label{H}
H(z(t))=\frac12\|p(t)\|^2+\frac12\sum_{i=1}^N\sum_{j\not= i}^NU(Q_{ij}(t)).
\end{equation}
\end{proposition}

\begin{proof}
The port-Hamiltonian formulation Eq.~(\ref{PHS}) reads
$$
\left[\begin{array}{c}\dot Q\\[1mm]\dot p \end{array}\right]=\left(\left[\begin{matrix}
    0& M\\
    -M^\text T& 0
    \end{matrix}\right]-\left[\begin{matrix}
    0&~0\\
    0&~\lambda I
    \end{matrix}\right]\right)\left[\begin{array}{c}\frac12\nabla U(Q)\\[1mm]p\end{array}\right]+\lambda\left[\begin{array}{c}0\\u\end{array}\right].
$$
The sparse structure of $M$ and the input  $u$ yields
$$
\dot Q = Mp,
$$
which implies 
$$
\dot Q_{ij}=p_i-p_j, 
$$
for each component.
The second equality is obtained by observing that
$$
\textstyle    \dot p=-M^\text T \frac12\nabla U(Q)-\lambda p +\lambda u.
$$
This equation reads for every $i=1,\dots,N$ as
$$
\begin{array}{lcl}
    \dot p_i&=&\lambda(u_i-p_i)-\frac12\sum_{j\not=i} \nabla U(Q_{ij})-\nabla U(Q_{ji}),\\[2mm]
    &=&\lambda(u_i-p_i)-\sum_{j\not=i} \nabla U(Q_{ij}),
    \end{array}
$$
where we used 
$\nabla U(Q_{ji})=-\nabla U(Q_{ij})$. 
Altogether, the microscopic model Eq.~(\ref{modmicro}) is recovered.
\end{proof}

\begin{figure}[!ht]
\centering\bigskip\small
\begin{minipage}[c]{.27\textwidth}
\centering\footnotesize Isotropic repulsive\\ potential (non-linear)\\[1mm]
\includegraphics[width=\textwidth]{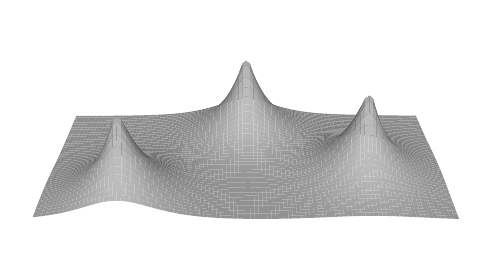}\\[2mm]
\includegraphics[width=.75\textwidth]{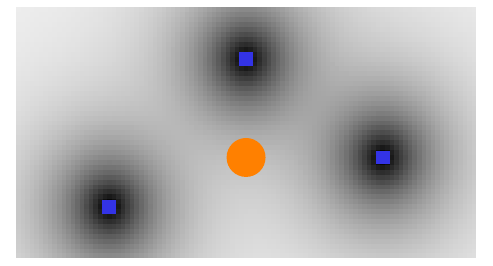}
\end{minipage}
~~~~~~~\textcolor{blue}{\huge$+$}~~~~~~~
\begin{minipage}[c]{.27\textwidth}
\centering Dissipation and\\ input control (linear)\\[1mm]
\includegraphics[width=\textwidth]{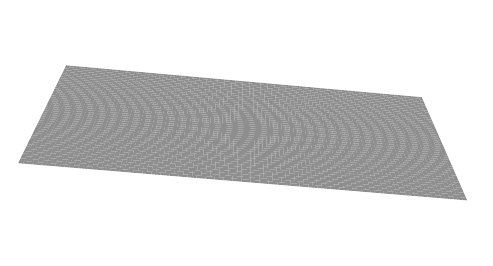}\\[1.2mm]
\includegraphics[width=.75\textwidth]{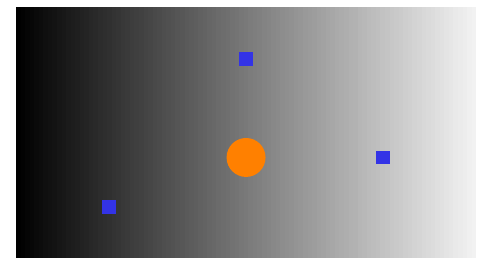}
\end{minipage}
~~~~~~~\textcolor{blue}{\huge$=$}~~~~~~~
\begin{minipage}[c]{.27\textwidth}
\centering Port-Hamiltonian\\ pedestrian model\\[1mm]
\includegraphics[width=\textwidth]{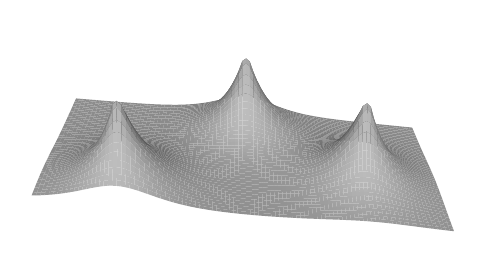}\\[1.5mm]
\includegraphics[width=.75\textwidth]{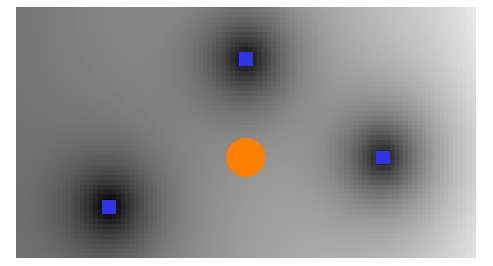}
\end{minipage}
\vspace{2mm}
\caption{Illustrative scheme for the port-Hamiltonian pedestrian model composed of non-linear isotropic repulsive interaction terms corresponding to the skew-symmetric structure of the PHS and linear dissipation and control ports.}
\label{figPHPM}
\end{figure}

\begin{remark}
In the port-Hamiltonian formulation, the isotropic structure of the interaction terms is crucial to obtain the skew-symmetric structure of the PHS.
Tactical decisions such as desired velocities are regarded as external information and provided through the input-port $u$. From a physical point of view, the tactical decisions adjust the energy of the system \cite{ortega2001putting}.
The parameter $\lambda$ accounts for the pedestrians' sensitivity and reactivity to the desired velocity. 
It appears in both the dissipation matrix and the input control port.
The result is a linear input-state-output PHS with dissipation and constant control input. 
The skew-symmetric part models the motion kinematic by pairwise repulsive social forces. This conservative part of the system yields the pedestrian proxemic and collision-avoidance behaviors. 
The dissipation matrix together with the input provides external information such as  destination or desired speed for each of the pedestrians.

In the presented formulation no interaction ports \cite[Eq.~(29)]{van2006port} are visible as the interaction ports of the individuals are already coupled in a PHS-preserving manner such that the system of all individual interactions admits a closed PHS structure.
Non-linear effects come from the isotropic interaction potential. 
The absence of boundary ports is because we consider the dynamics on the torus.
We illustrate the different components of the port-Hamiltonian pedestrian model in Fig.~\ref{figPHPM}.
Here the orange point is the considered pedestrian, while the three blue squares are neighbors.
The grey levels in the background quantify the acceleration, the grey level being lighter as the acceleration increases. 
The plots on the left represent the non-linear isotropic repulsive interaction terms corresponding to the skew-symmetric structure of the PHS. 
The central plots describe the linear dissipation and control ports, while the plots on the right, summing these two components, represent the resulting acceleration field of the port-Hamiltonian pedestrian model.
\end{remark}

\begin{remark}
Our Hamiltonian formulation is based on isotropic and distance--dependent interactions.
In the literature, most force-based pedestrian models include anisotropic repulsion mechanisms to account for fundamental diagrams \cite{seyfried2005fundamental}, 
vision cone effects \cite{Helbing1995,chraibi2010generalized}, 
or preferred crossing directions \cite{Totzeck2020}. 
The repulsion terms are weighted by factors generally depending on the bearing angle, i.e., the velocity, breaking the skew--symmetry of the Hamiltonian formulation. 
In the port-Hamiltonian framework, anisotropic effects can be modelled for example with the help of the state--dependent input terms. Some possible extensions of the PHS model, including anisotropic effects, are discussed in the conclusion.
\end{remark}

\subsection{Hamiltonian behavior}
Exploiting the PHS structure directly we obtain the energy balance
\begin{equation}\label{balance}
\frac d{dt}H( z(t))= y^\text T(z(t)) \tilde u(t)- \nabla^\text T\! H(z(t))\, R\,\nabla H(z(t))=\lambda\langle p(t),u(t)-p(t) \rangle.
\end{equation}
Hence the time derivative of the Hamiltonian depends only on the pedestrians velocities. We observe that the time derivative is zero if all pedestrians move with their desired velocities, i.e.,
\begin{equation}
\frac d{dt}H( z(t))=0\quad\text{if}\quad \forall i,~ p_i(t)=u_i(t).
\end{equation}
Moreover, the following simple implications hold:
\begin{itemize}
    \item The Hamiltonian is constant, $\frac d{dt}H(z(t))=0$ for all $t\ge0$, if the system has no dissipation, i.e., 
    \begin{equation}\label{dHlambda=0}
    \forall t\ge0,~ \frac d{dt}H( z(t))=0\quad\text{if}\quad \lambda=0.
    \end{equation}
    Indeed, the purely Hamiltonian systems is conservative as expected.
   
    \item If $u\equiv 0$, the Hamiltonian decreases over time,
    \begin{equation}\label{dHu=0}
    \forall t\ge0,~ \frac d{dt}H( z(t))\le0\quad\text{if}\quad \forall i,~ u_i=0.
    \end{equation}
    Dissipation yields asymptotic stability with crystallisation as equilibrium.
    \item The Hamiltonian is also decreasing if $p$ and $u$ are orthogonal. More generally, the Cauchy–Schwarz inequality provides
\begin{equation}
\langle p(t),u(t)\rangle-\|p(t)\|^2\le\|p(t)\|\|u(t)\| -\|p(t)\|^2.
\end{equation}
Therefore 
\begin{equation}\label{dHneg}
\frac d{dt}H(z(t))\le\lambda\|p(t)\|(\|u(t)\|-\|p(t)\|)\le0, \quad \text{if} \quad \|u(t)\|\le\|p(t)\|.
\end{equation}
\end{itemize}
However, there is no monotone relationship between $H$ and $\|p\|$. The speed may increase even if the Hamiltonian decreases, if, for instance, the distances between the individuals increases. Therefore, having initially $\|u\|\le\|p^0\|$ is not sufficient to demonstrate that $\frac d{dt}H(z(t))\le0$ for all $t\ge0$.

\begin{remark}
A final remark concerns the asymptotic behaviour of the Hamiltonian without pedestrian interaction.
Assuming the desired velocities $u$ to be constant over time and no interaction takes place, i.e., $A=0$, the pedestrian velocities simply relax to the desired velocity and the asymptotic behaviour characterized by
\begin{equation}\label{dHA=0}
    H(t)\rightarrow H^\ast(u) :=\frac12\|u\|^2,\quad\text{as}\quad t\rightarrow\infty. 
\end{equation} 
This estimate allows to provide a physical threshold of interaction for the Hamiltonian that lays the ground for the order parameter that will be discussed in detail in Section~\ref{simulation}. 
\end{remark}

\newpage

\section{Numerical schemes and simulation results}

\subsection{Comparison of numerical schemes}
In this section, we test different numerical schemes for the simulation of the particle system. 
We denote the acceleration of the $i$-th pedestrian as
\begin{equation}
    a(q_i,p_i)=\lambda\big(u_i-p_i\big)-\sum_{j\not=i}\nabla U\big(q_i-q_j\big).
\end{equation}
The tested schemes are combinations of explicit (forward) and implicit (backward) Euler methods applied to particle speeds and positions and a truncated version of the leapfrog algorithm.
\begin{itemize}
\item Euler explicit/explicit scheme:
\begin{equation}\begin{cases}
~p^{k+1}=p^k+\dt \; a(q^k,p^k)\\[1mm]
~q^{k+1}=q^k+\dt \; p^{k}
\end{cases}\label{EulerEE}\end{equation}

\item Euler explicit/implicit scheme:
\begin{equation}\begin{cases}
~p^{k+1}=p^k+\dt \; a(q^k,p^k)\\[1mm]
~q^{k+1}=q^k+\dt \; p^{k+1}
\end{cases}\label{EulerEI}\end{equation}

\item Euler implicit/explicit scheme:
\begin{equation}\begin{cases}
\displaystyle ~p^{k+1}=p^k+\dt \; a(q^{k+1},p^{k+1})=p^k+\frac{\dt }{1+\lambda \dt }a(q^{k+1},p^k)\\[1mm]
~q^{k+1}=q^k+\dt \; p^{k}
\end{cases}\label{EulerIE}\end{equation}

\item Euler implicit/implicit scheme: 
\begin{equation}\begin{cases}
~p^{k+1}=p^k+\dt \; a(q^{k+1},p^{k+1})\\[1mm]
~q^{k+1}=q^k+\dt \; p^{k+1}
\end{cases}\label{EulerII}\end{equation}
Here the equation $p^{k+1}=p^k+\dt \; a(q^{k+1},p^{k+1})$  has to be solved numerically.

\item Leapfrog scheme:
$$\begin{cases}
~\displaystyle p^{k+1}=p^k+\frac{\dt }2\big(a(q^k,p^k)+a(q^{k+1},p^{k+1})\big)\\[2mm]
~\displaystyle q^{k+1}=q^k+\dt \;p^k+\frac{\dt ^2}2 a(q^k,p^k)
\end{cases}$$
Note that 
$$
a(q^{k+1},p^{k+1})
=a(q^{k+1},p^k)+\lambda(p^k-p^{k+1}),
$$
and the leapfrog scheme reads after simplification
\begin{equation}\begin{cases}
~\displaystyle p^{k+1}=p^k+\frac{\dt }{2+\lambda \dt }\big(a(q^k,p^k)+a(q^{k+1},p^k)\big)\\[2mm]
~\displaystyle q^{k+1}=q^k+\dt \;p^k+\frac{\dt ^2}2 a(q^k,p^k)
\end{cases}\label{leapfrog}\end{equation}
\end{itemize}

\begin{figure}[!ht]
\centering\medskip
\includegraphics[width=8.7cm]{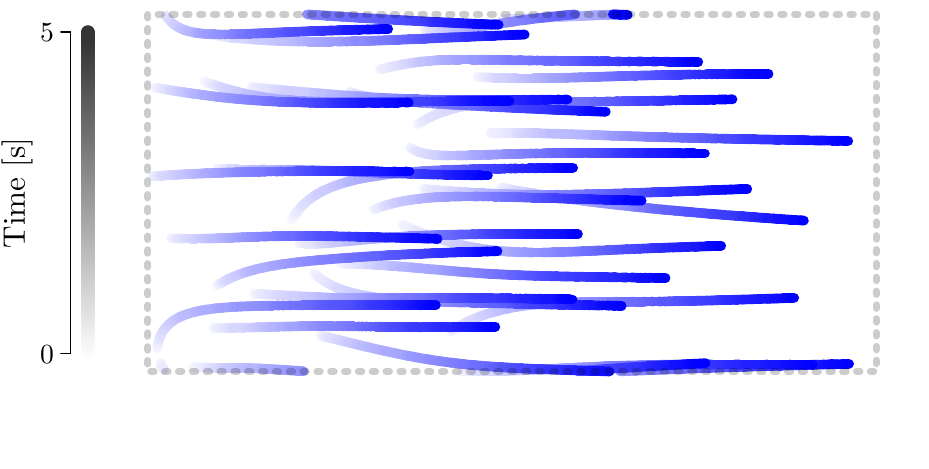}\footnotesize\input{Figures/H1s.tex}\vspace{-2mm}
\caption{Simulation of a uni-direction flow used as a reference to compare the six numerical schemes Eqs.~(\protect\ref{EulerEE})-- (\protect\ref{leapfrog}), see Fig.~\protect\ref{fig:Error}. Illustrative example of the leapfrog scheme (\protect\ref{leapfrog}) with $\dt=0.001$~s.}
\label{fig:TrajUD}\end{figure}

\begin{figure}[!ht]
\centering\bigskip\medskip
\input{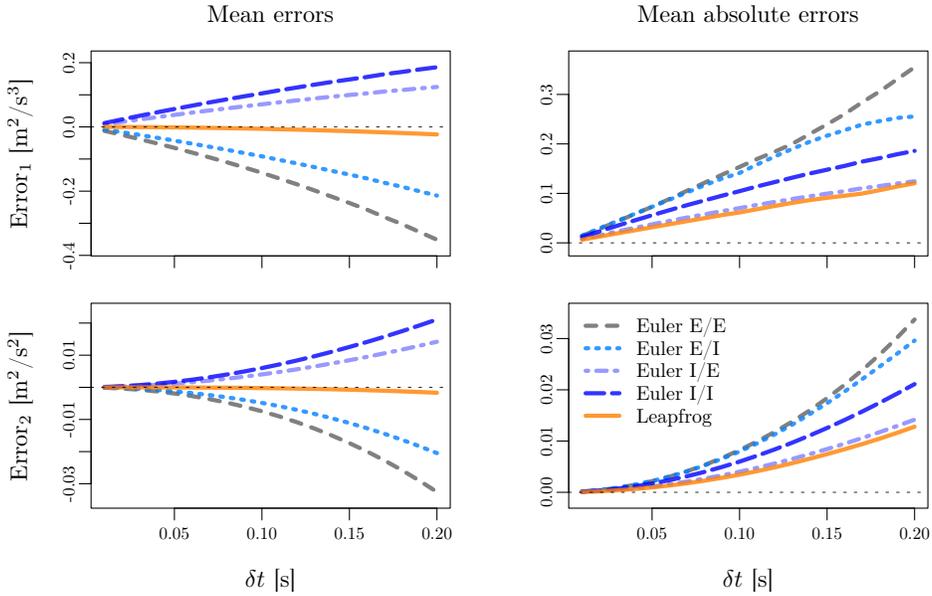}\vspace{-2mm}
\caption{Mean numerical errors of the Euler and leapfrog discretisation schemes averaged during the 20 first seconds of the experiment in Fig.~\protect\ref{fig:TrajUD} for numerical time steps $\delta t$ ranging from 0.01 to 0.2~s. Top panels: Error resulting from the numerical approximation of the time derivative of the Hamiltonian. Bottom panels: Error resulting from the numerical approximation of the Hamiltonian. The left panels show the mean errors while the right panels show the mean absolute errors. The leapfrog scheme systematically outperforms the implicit and explicit Euler schemes.} 
\label{fig:Error}
\end{figure}

We perform a simulation experiment to compare the precision of the different numerical schemes. 
We simulate the evolution of 32 particles on an $11$ meters $\times$ $5$ meters domain with periodic boundary conditions, which can be interpreted as torus. 
The desired velocity $u = (1,0)^\top$ is pointing to the right. 
Initially the particles are randomly distributed on the left part of the domain with zero velocity. 
Results of an exemplary simulation are presented in Fig.~\ref{fig:TrajUD}.
The parameters values are chosen as in standard models \cite{Helbing1995,tordeux2016collision}, see Table~\ref{tab:StdPara} for more details. 
To quantify whether the port-Hamiltonian dynamics are accurately described, we compare the time derivative of the Hamiltonian provided by the balance equation (\ref{balance}) to the time-difference of the Hamiltonian Eq.~(\ref{H}),
\begin{equation}
    \text{Error}_1^k=\lambda\langle p^k,u-p^k \rangle-\frac1\dt\left(H(Q^k,p^k)-H(Q^{k-1},p^{k-1})\right).
\end{equation}
The errors should vanish as $\delta t\rightarrow0.$
We also compute the numerical integral
\begin{equation}
    \text{Error}_2^k=\dt\sum_{i=1}^k\text{Error}_1^k=H(Q^0,p^0)+\lambda\dt\sum_{i=0}^k\langle p^i,u-p^i\rangle-H(Q^k,p^k),
\end{equation}
to directly assess the Hamiltonian estimation precision.  
Such errors, averaged over the first 20 seconds of the simulation, are presented in Fig.~\ref{fig:Error} for the five schemes and numerical time steps $\delta t$ ranging from 0.01 to 0.2~s. 

\begin{figure}[!ht]
\centering\bigskip\medskip
\input{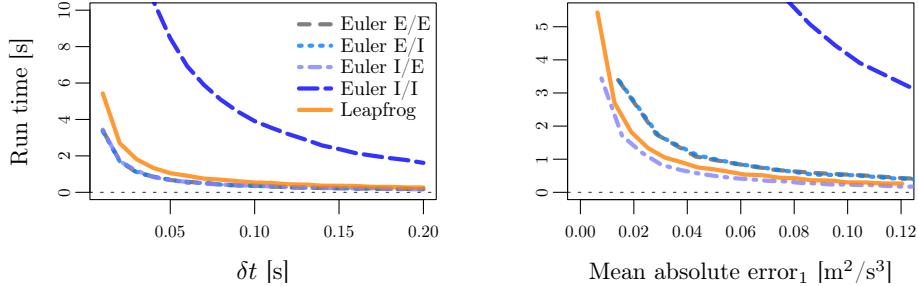}\vspace{-2mm}
\caption{Run times of the Euler and leapfrog discretisation schemes for the simulation of the 20 first seconds of experiment in Fig.~\protect\ref{fig:TrajUD} according to the numerical time step $\delta t$ (left panel) and the numerical absolute error (right panel). In terms of run time, the explicit Euler schemes are faster than the leapfrog scheme (left panel). In terms of CPU resources, the implicit/explicit Euler scheme provides the best approximation, followed by the leapfrog scheme, the other explicit Euler schemes, and, well afterwards, the implicit Euler scheme (right panel).} 
\label{fig:Error2}
\end{figure}

Two main characteristics are noteworthy:
\begin{itemize}
    \item The explicit schemes tend to underestimate the Hamiltonian and its derivative (positive bias) while the implicit schemes overestimate it (negative bias, see Fig.~\ref{fig:Error}, left panels). 
    The leapfrog algorithm error is much more centered and presents almost unbiased estimates. 
    \item In terms of absolute error, the explicit/explicit and explicit/implicit Euler schemes have similar behaviors as do the implicit/explicit Euler and leapfrog schemes (see Fig.~\ref{fig:Error}, right panels); the second pair being much more accurate. 
    The implicit/implicit Euler scheme provides intermediate estimates. 
\end{itemize}
These observations are in accordance with well-known results from ODE theory.
Figure \ref{fig:Error2} presents the scheme efficiency in terms of CPU time. 
Here, we measure the time required by the different schemes to simulate 20 seconds of the experiment shown in Fig.~\protect\ref{fig:TrajUD}. 
The three explicit Euler schemes Eqs.~(\ref{EulerEE}), (\ref{EulerEI}) and (\ref{EulerIE}) need similar resources. 
The leapfrog scheme Eq.~(\ref{leapfrog}) requires two computations of the acceleration function at each time step and is therefore less effective (by approximately a factor 1.6, see Fig.~\ref{fig:Error2}, left panel). Even more computational effort (up to a factor 10) is needed by the implicit Euler scheme Eq.~(\ref{EulerII}) which uses several computations of the acceleration function. 
On the other hand, the implicit/explicit Euler scheme provides the best approximation in terms of CPU resources for any precision error (see Fig.~\ref{fig:Error2}, right panel). 
Followed by the leapfrog scheme, the other explicit Euler schemes, and, well afterwards, the implicit Euler scheme. 
Simulation of other scenarios, including counter and crossing flows, showed similar behaviors. Again, the observations coincide with the general theory. The following results are based on the leapfrog discretization.

\begin{table}[!ht]
\renewcommand{\arraystretch}{1.5}
    \centering
    \begin{tabular}{l|l|l|ll}
        Parameter&Interpretation&PHS component&Value\\
        \hline
        $\lambda$&Pedestrian sensitivity&Dissipation matrix&$2$\,s$^{-1}$\\
        $|u|$&Desired speed&Input control port&$1$\,m/s\\
        $A$&Repulsion strengh&Skew-symmetric structure&$5$\,ms$^{-2}$\\
        $B$&Interaction range&Skew-symmetric structure&$0.3$\,m\\
    \end{tabular}\bigskip
    \caption{Model parameters and values used for simulation.}
    \label{tab:StdPara}
\end{table}

\newcommand{\titleFigTraj}[1]{\bigskip\hspace{-5cm}{#1}\\[-0mm]}
\newcommand{\captionFigTraj}[4]{Simulation of a #1 with the port-Hamiltonian pedestrian model\if#31 for #2\fi. Leapfrog scheme (\protect\ref{leapfrog}) with $\dt=1\text{e-}3$~s. #4}

\subsection{Simulation results}\label{simulation}

The simulation results presented in the following describe the evolution of 32 particles on an $11$ meters $\times$ $5$ meters domain with periodic boundary conditions. 
We consider a mixed flow with opposite desired directions.
The blue particles systematically endeavor to move to the right. 
The orange particles aim to move to the left for the counter-flow experiment, while the desired direction is to the top for the crossing-flow. 
The blue and orange particles are initially randomly distributed on the left and right sides of the domain, respectively, in the counter-flow experiment. 
They are indifferently initially randomly distributed over the domain for the crossing-flow experiment.
The initial velocities of all particles are zero for both experiments. 
The model parameter settings are given in Table~\ref{tab:StdPara}. 
All simulations are carried out on NetLogo\footnote{An NetLogo online simulation platform of the port-Hamiltonian pedestrian model is available at
\href{https://www.vzu.uni-wuppertal.de/fileadmin/site/vzu/Port-Hamiltonian_pedestrian_model.html?speed=0.5}{\texttt{https://www.vzu.uni-wuppertal.de/fileadmin/site/vzu/Port-Hamiltonian\_pedestrian\_model}}.} \cite{wilensky1999netlogo} using the leapfrog scheme (\protect\ref{leapfrog}) with time step $\dt=0.001$\;s.

\subsubsection{Basic dynamics}
\begin{figure}[!ht]
\titleFigTraj{{\hspace{-2cm}\small\bf Dynamics without dissipation} ~($\lambda=0$)}\centering
\includegraphics[width=8.7cm]{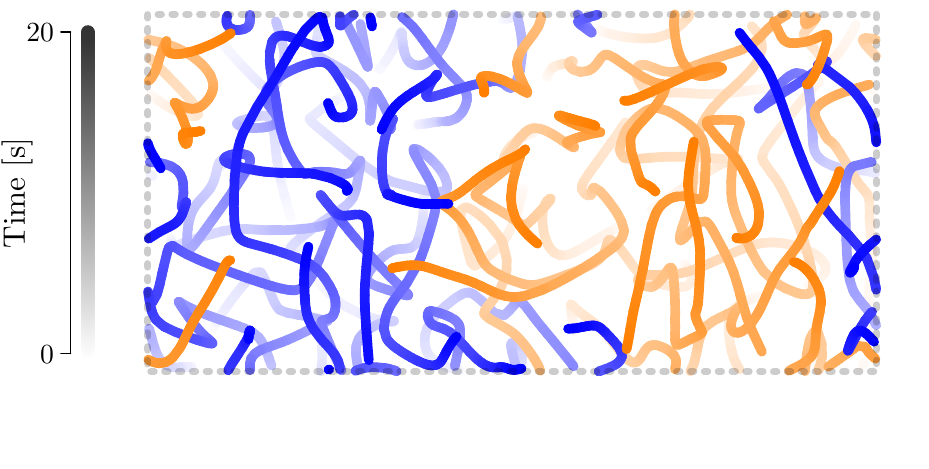}\footnotesize\input{Figures/H2s.tex}\vspace{-2mm}
\caption{\captionFigTraj{counter flow}{$\lambda=0$ (pure Hamiltonian case)}{1}{The total energy is conserved, the Hamiltonian is constant and the particle dynamics are close to those of a dynamical billiard (2D colloid).}}
\label{fig:Trajl0}\end{figure}

\begin{figure}[!ht]
\titleFigTraj{{\hspace{3cm}\small\bf Dynamics with dissipation and without input control} ~($\lambda>0, \,u=0$)}\centering
\includegraphics[width=8.7cm]{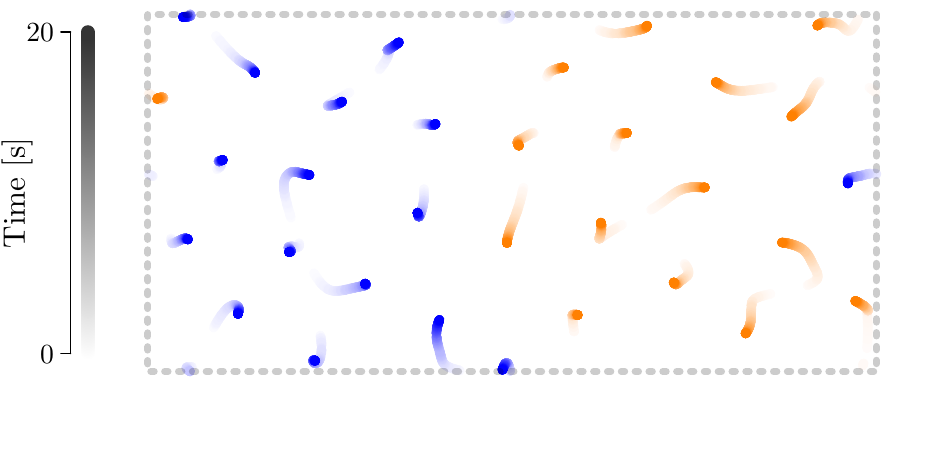}\footnotesize\input{Figures/H4s.tex}\vspace{-2mm}
\caption{\captionFigTraj{counter flow}{$u=0$ (dissipative Hamiltonian case)}{1}{The energy in the system is progressively dissipated, and the particles crystallise to a homogeneous equilibrium configuration. The Hamiltonian relaxes towards low energy levels.}}
\label{fig:Traju0}\end{figure}

\begin{figure}[!ht]
\titleFigTraj{{\small\bf Dynamics with no pedestrian interaction} ~($A=0$)}\centering
\includegraphics[width=8.7cm]{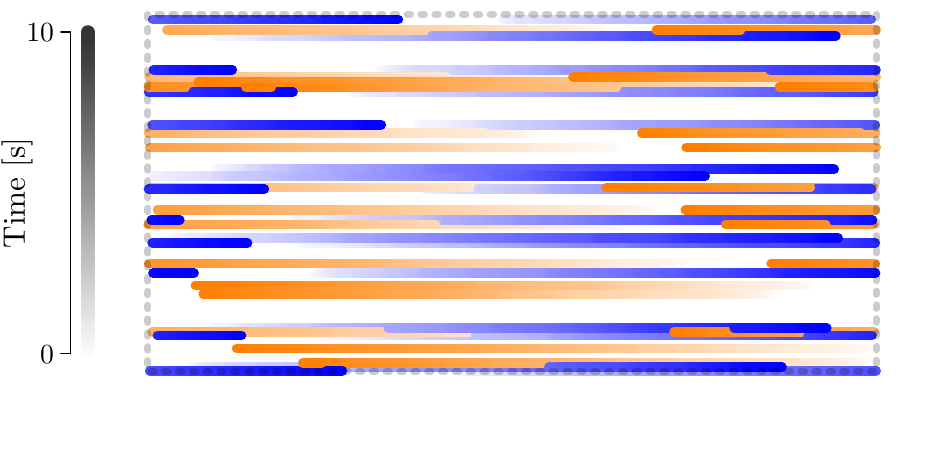}\footnotesize\input{Figures/H3s.tex}\vspace{-2mm}
\caption{\captionFigTraj{counter flow}{$A=0$ (no interaction)}{1}{The particles relax their velocities towards the desired $u$ while the Hamiltonian converges to $H^\star=Nu^2/2=16$.}}
\label{fig:TrajA0}\end{figure}

The following simulation results show that the port-Hamiltonian pedestrian model can describe different fundamental dynamics. 
Let us discuss the basic dynamics:
\begin{itemize}
    \item  Case $\lambda=0$. No dissipation to the desired velocity occurs if the relaxation rate $\lambda$ is zero. The system is purely Hamiltonian and the total energy is conserved, see Eq.~(\ref{dHlambda=0}). For $A$ large the system describes disordered dynamics similar to dynamical billiards (2D colloids) governed by the repulsion potential between the particles, see Fig.~\ref{fig:Trajl0}.
    \item Case $u=0$.~ Adding dissipation (i.e., $\lambda>0$) and neglecting the input control, i.e.~$u\equiv 0$ results in crystallization phenomena (see Fig.~\ref{fig:Traju0}). Indeed, the particles have no preferred direction and the desired velocity is zero. The particles may even reach deterministic hexagonal or squared homogeneous grids according to the interaction distance $D$ if the relaxation $\lambda$ is sufficiently low. The derivative of the Hamiltonian is  nonpositive, see Eq.~(\ref{dHu=0}). Hence, the system is dissipative and the Hamiltonian relaxes towards low energy levels.
    \item Case $A=0$.~ No interactions arise when the interaction potential is zero. The agents relax their speeds to the desired velocity and the Hamiltonian trivially relaxes towards $H^\star=Nu^2/2,$ see Eq.~(\ref{dHA=0}) and Fig.~\ref{fig:TrajA0}.
\end{itemize}

\subsubsection{Collective dynamics}

\begin{figure}[!ht]
\titleFigTraj{{\small\bf Lane formation for a counter flow} ~($\lambda=2$ s$^{-1}$)}\centering
\includegraphics[width=8.7cm]{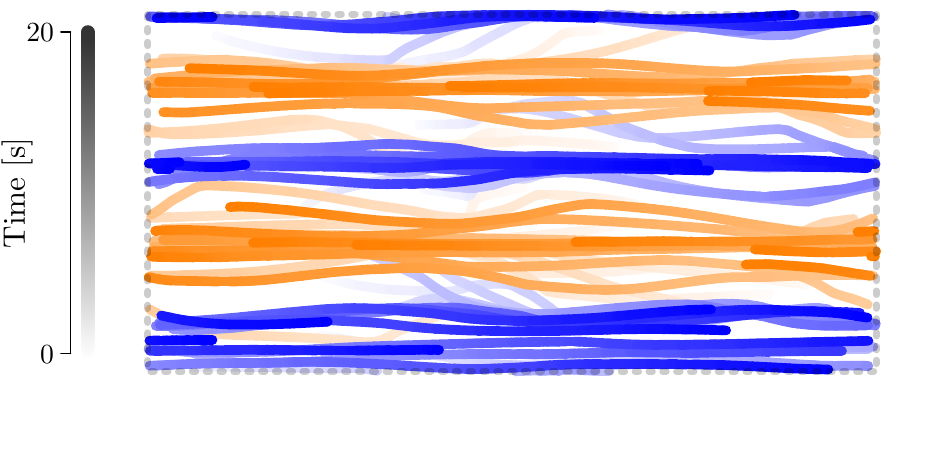}\footnotesize\input{Figures/H5s.tex}\vspace{-2.5mm}
\caption{\captionFigTraj{counter flow}{$\lambda=2$ s$^{-1}$ (reactive pedestrian)}{1}{Lane formation by motion direction quickly emerges in the dynamics. The Hamiltonian initially fluctuates before stabilising to a value higher than $H^\ast$.}}
\label{fig:TrajLF}\end{figure}

\begin{figure}[!ht]
\titleFigTraj{{\hspace{-1cm}\small\bf Gridlock for a counter flow} ~($\lambda=0.1$ s$^{-1}$)}\centering
\includegraphics[width=8.7cm]{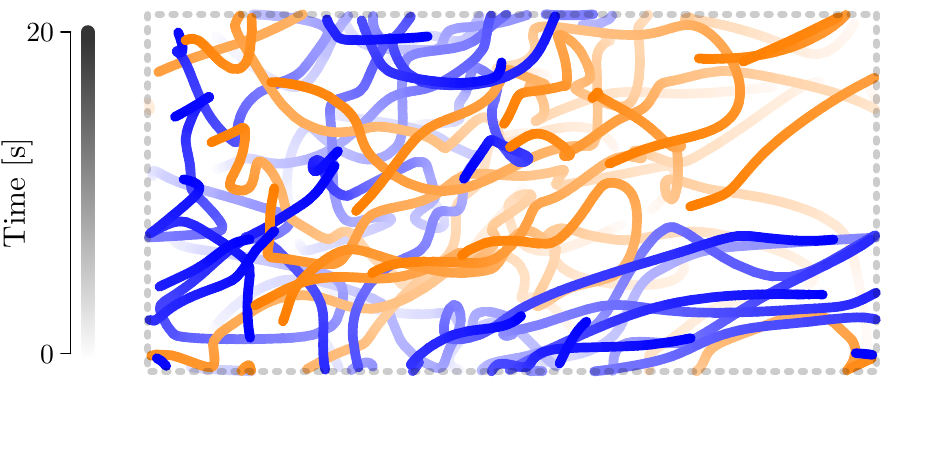}\footnotesize\input{Figures/H6s.tex}\vspace{-2.5mm}
\caption{\captionFigTraj{counter flow}{$\lambda=0.1$ s$^{-1}$ (low pedestrian reactivity)}{1}{No lane formation emerges when the pedestrians are not sufficiently reactive, yielding in gridlocks in the dynamics. The Hamiltonian presents large fluctuations and remains lower than $H^\ast$.}}
\label{fig:TrajLFb}\end{figure}

\begin{figure}[!ht]
\titleFigTraj{{\small\bf Strip formation for a crossing flow} ~($\lambda=2$ s$^{-1}$)}\centering
\includegraphics[width=8.7cm]{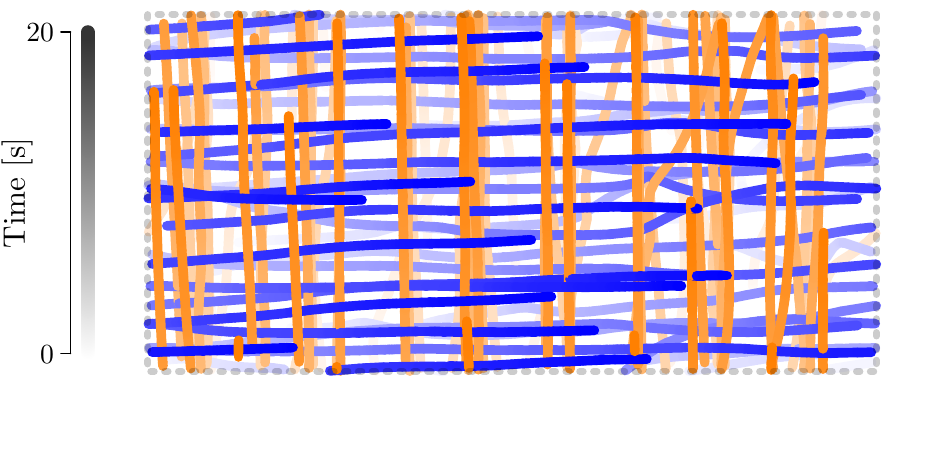}\footnotesize\input{Figures/H7s.tex}\vspace{-2.5mm}
\caption{\captionFigTraj{counter flow}{$\lambda=2$ s$^{-1}$ (reactive pedestrian)}{1}{Strip formation diagonal to the  motion direction quickly emerges in the dynamics. The configuration for long simulation times is arranged in such a way that pedestrians pass each other without deviating from their desired direction. The Hamiltonian stabilises to a value higher than $H^\ast$.}}
\label{fig:TrajSF}\end{figure}

\begin{figure}[!ht]
\titleFigTraj{{\small\bf Partial gridlock for a crossing flow} ~($\lambda=0.1$ s$^{-1}$)}\centering
\includegraphics[width=8.7cm]{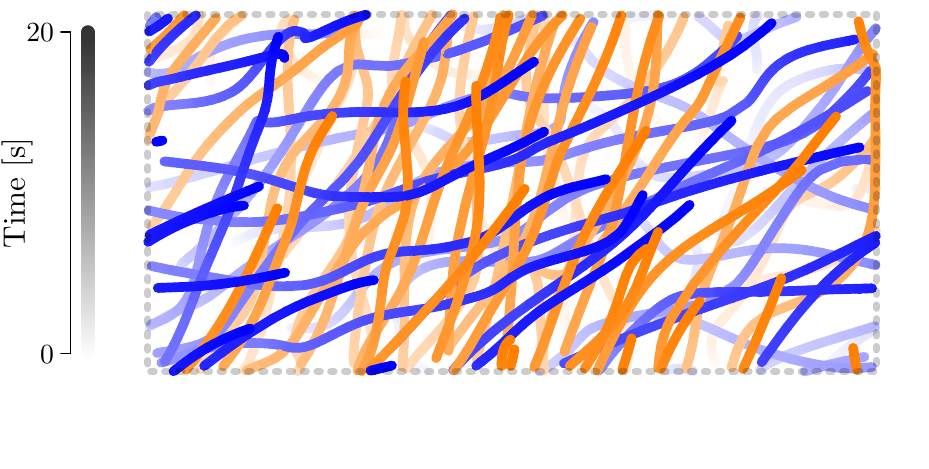}\footnotesize\input{Figures/H8s.tex}\vspace{-2.5mm}
\caption{\captionFigTraj{counter flow}{$\lambda=0.1$ s$^{-1}$ (low pedestrian reactivity)}{1}{Strip formation also partly emerges in the dynamics. However, in contrast to the strips obtained for high pedestrian reactivity (see Fig.\protect\ref{fig:TrajSF}), the strips are partly opposed to the desired motion direction, yielding in the formation of diagonal gridlocks deviating the pedestrians. The Hamiltonian remains lower than $H^\ast$.}}
\label{fig:TrajSFb}\end{figure}

The previous studies illustrated the different model components and their roles in the dynamics.
However, none of these is realistic for pedestrian dynamics. 
Force-based models can describe many types of collective dynamics observed in pedestrian crowds, including lane and strip formation for counter and crossing flows, or jamming, arching, and alternating counter flow at bottlenecks \cite{Helbing1995,Boltes2018,Schadschneider2018}. 
A phase transition occurs from disordered states to coordinated dynamics. 
Critical parameter settings partition the phase diagram of the system.
In the model, lane or strip formation arises for reactive agents, i.e., for $\lambda$ sufficiently large, while gridlocks occur as the relaxation operates slowly (see Figs.~\ref{fig:TrajLF} and \ref{fig:TrajLFb} for counter flow and the lane formation, or Figs.~\ref{fig:TrajSF} and \ref{fig:TrajSFb} for counter flow and the strip formation). 
Numerical tests indicate that the dynamics converge to stable stationary states in case of collective dynamics (i.e., $\lambda$ large) while they are unstable and seem not to converge for $\lambda$ small. 
Indeed, in the port-Hamiltonian framework, the parameter $\lambda$ balances the dynamics between the skew-symmetric parts enforcing conservation of energy and the dissipation and control input components which vary the energy and lead to long-term stability. 
Note that other parameters, for instance, the interaction range $B$, also significantly influence the dynamics. Order parameters are helpful tools to distinguish the different behaviours discussed here. In the following section, we introduce an order parameter that is based on the Hamiltonian. Here, the advantage of the PHS structure of the proposed model become apparent.

\section{Hamiltonian as physical order parameter}

One challenge in multiscale modelling is the formulation of generic macroscopic order parameters that allow to quantify the system state. 
Many order parameters have been developed in the literature.
In most cases, the order parameters are tailored to detect specific types of collective dynamics. 
For instance, an order parameter quantifying lane formation reads 
\begin{equation}\label{OPL}
\Phi_L=\frac1N\sum_{i=1}^N\phi_i^L
\end{equation}
with
$$
\phi_i^L=\Big[\frac{L_i-\underline L_i}{L_i+\underline L_i}\Big]^2\qquad\text{and}\quad
\left.\begin{array}{l}
L_i=\text{card}\big(j,~|y_i-y_j|<\Delta,~u_i=u_j\big),\\[1mm]
\underline L_i=\text{card}\big(j,~|y_i-y_j|<\Delta,~u_i\ne u_j\big),\end{array}\right.
$$
and $(x_i,y_i)_i$ are the positions of the agents and $(u_i)_i$ their desired velocities. 
Here $\Delta=0.5$~m, $2\Delta$ being a lane width threshold value.
The parameter (\ref{OPL}) was initially introduced to detect lanes in a colloidal suspension \cite{Rex2007} and used in pedestrian dynamics as well \cite{Nowak2012,khelfa2022heterogeneity}.
Similarly, an order parameter to detect diagonal strip formation may be formulated as
\begin{equation}\label{OPS}
\Phi_S=\frac1N\sum_{i=1}^N\phi_i^S,
\end{equation}
with
$$
\phi_i^S=\Big[\frac{S_i-\underline S_i}{S_i+\underline S_i}\Big]^2\qquad\text{and}\quad
\left.\begin{array}{l}
S_i=\text{card}\big(j,~|y_i-y_j+x_i-x_j|<\Delta,~u_i=u_j\big),\\[1mm]
\underline S_i=\text{card}\big(j,~|y_i-y_j+x_i-x_j|<\Delta,~u_i\ne u_j\big).\end{array}\right.
$$ 
The values of both order parameters, $\phi_i^L$ and $\phi_i^S$, range from zero, when the directions of pedestrians of the same lane or band are uniformly mixed, to one when the pedestrians walk in the same direction.

The Hamiltonian measures the energy in the system.
Its value remarkably well quantifies the collective dynamics in the pedestrian port-Hamiltonian model. 
Indeed, it is systematically high and stable when collective dynamics occur (see Figs.~\ref{fig:TrajLF} and \ref{fig:TrajSF}, top right panel), while it is low and fluctuating for disordered states  (see Figs.~\ref{fig:TrajLFb} and \ref{fig:TrajSFb}). 
The threshold $H^\ast$ corresponds to the Hamiltonian for agents moving at desired velocities with no interaction, see Eq.~\eqref{dHA=0}. 
We observe that the Hamiltonian is higher than $H^\ast$ in case of collective behaviors. 
Indeed, collective dynamics are motions at desired velocity in interaction with neighboring agents. 
The repulsive interaction potentials lead to values larger than $H^\ast$. 
On the other hand, the Hamiltonian is lower than $H^\ast$ in case of disordered dynamics with low speed terms and fluctuating distance-based potential terms. 
Based on these observations we propose an order parameter based on the Hamiltonian given by 
\begin{equation}\label{OPH}
    \Phi_H=\frac1{1+\exp(\kappa(H^\ast-H))}, \qquad\kappa=100.
\end{equation} 
With this choice we aim to emphasize the difference between the Hamiltonian $H$ and the Hamiltonian value $H^\ast$ at equilibrium without interaction, see Eq.~\eqref{dHA=0}. 

\begin{figure}[!ht]
\centering\bigskip
\input{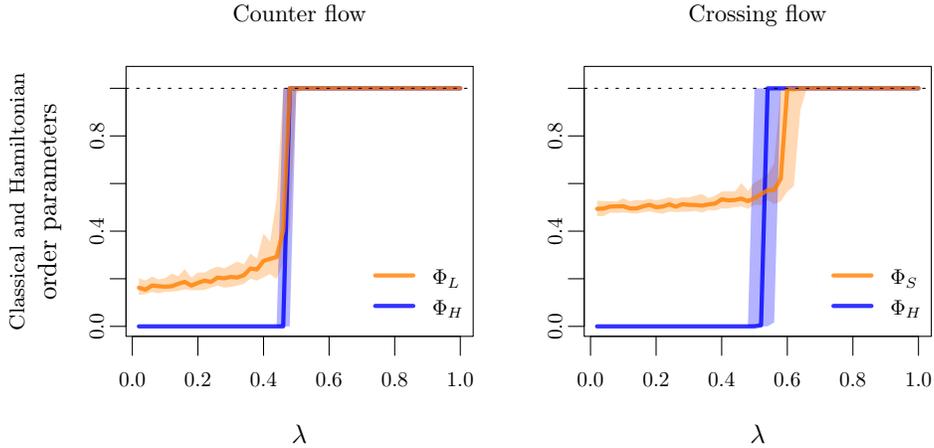}\vspace{-4mm}
\caption{Classical order parameters for lane and band formation Eqs.~(\protect\ref{OPL}) and (\protect\ref{OPS}) and Hamiltonian order parameter Eq.~(\protect\ref{OPH}) for the counter and crossing flow experiments. The curves are median values over 1000 independent simulations while the band are inter-quartile range. Both parameters describe similar phase transitions. Yet, in contrast to classical order parameters, the Hamiltonian parameter is generic and not tailored for the specific collective dynamics at hand.} 
\label{fig:OP}
\end{figure}

A comparison to classical order parameters for lane and strip formation Eqs.~(\ref{OPL}) and (\ref{OPS}), respectively, is shown in Fig.~\ref{fig:OP}.
The Hamiltonian order parameter $\Phi_H$ polarises in zero and one as $H<H^\ast$ and $H>H^\ast$, respectively.
The classical and Hamiltonian order parameters are measured for long simulation times with $\lambda$ ranging from 0.01 to 1~s$^{-1}$; the dynamics being disordered for low $\lambda$ (see Figs.~\ref{fig:TrajLFb} and \ref{fig:TrajSFb}) and present lanes or strips for large $\lambda$ (see Figs.~\ref{fig:TrajLF} and \ref{fig:TrajSF}). 
Interestingly, the critical values for which the classical and Hamiltonian order parameters present a transition coincide for both, counter and crossing flow experiments, see Fig.~\ref{fig:OP}. 
In contrast to classical order parameters, the Hamiltonian parameter is generic and not tailored for the specific collective dynamics at hand.

\section{Summary and model development perspectives}\label{conclusion}

In this article, we identify a class of force--based pedestrian models that can be formulated as port--Hamiltonian systems.
The general class relies on two main components: 
\begin{itemize}
    \item Isotropic distance-based repulsion between the pedestrians appearing in the skew-symmetric part of the Hamiltonian system.
    \item Relaxation to the desired velocity provided exogenously from a tactical modelling level which corresponds to dissipation and control. 
\end{itemize}
The resulting PHS is a linear input-state-output port-Hamiltonian system with no interaction port \cite{van2006port}. 
Non-linear effects come from the isotropic interaction potential. 
Despite its simplicity, the pedestrian model exhibit many types of collective dynamics of force-based models, e.g., lane formation for counter-flow or strip formation for crossing-flow. 
Interestingly, the Hamiltonian describes systematic behaviors according to the dynamics, being high and stable in case of collective motion and low and fluctuating in case of disorder. 
A critical Hamiltonian threshold value relying only on the input control could be identified and physically interpreted.
Using the Hamiltonian as a generic macroscopic order parameter in a port-Hamiltonian framework is a natural and promising multiscale modelling approach for pedestrian dynamics. Moreover, there is no need to adapt the order parameter for the specific dynamics at hand.

\begin{figure}[!ht]
\centering\bigskip\bigskip
\begin{minipage}[c]{.21\textwidth}
\centering Isotropic repul- sive interaction\\[1mm]
\includegraphics[width=\textwidth]{Figures/sk11.pdf}
\end{minipage}
\textcolor{blue}{\LARGE$\times$}
\begin{minipage}[c]{.21\textwidth}
\centering Anisotropic effect\\[1mm]
\includegraphics[width=\textwidth]{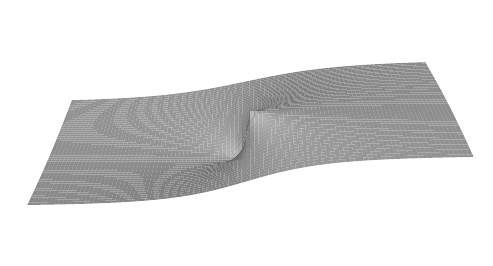}
\end{minipage}
\textcolor{blue}{\LARGE$+$}
\begin{minipage}[c]{.21\textwidth}
\centering Dissipation and\\ input control\\[1mm]
\includegraphics[width=\textwidth]{Figures/sk31.pdf}
\end{minipage}
\textcolor{blue}{\LARGE$=$}
\begin{minipage}[c]{.21\textwidth}
\centering Force-based\\model\\[1mm]
\includegraphics[width=\textwidth]{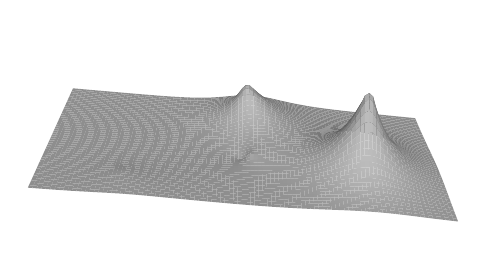}
\end{minipage}\\[7mm]

\begin{minipage}[c]{.21\textwidth}
\centering Isotropic repul- sive interaction\\[1mm]
\includegraphics[width=\textwidth]{Figures/sk11.pdf}
\end{minipage}
\textcolor{blue}{\LARGE$+$}
\begin{minipage}[c]{.21\textwidth}
\centering Anisotropic effect\\[1mm]
\includegraphics[width=\textwidth]{Figures/sk21.pdf}
\end{minipage}
\textcolor{blue}{\huge$?$}
\begin{minipage}[c]{.21\textwidth}
\centering Dissipation and\\ input control\\[1mm]
\includegraphics[width=\textwidth]{Figures/sk31.pdf}
\end{minipage}
\textcolor{blue}{\LARGE$=$}
\begin{minipage}[c]{.21\textwidth}
\vspace{-5mm}
\centering {Anisotropic port-Hamiltonian pedestrian\\ model}
\end{minipage}
\medskip
\caption{Illustrative scheme for possible future development of the port-Hamiltonian pedestrian model incorporating anisotropic effects (e.g., vision field). Top panels: classical force-based framework. Bottom panels: desired port-Hamiltonian formulation.}
\label{figPHPMb}
\end{figure}

The Hamiltonian formulation exploits that the pedestrian interactions are assumed to be isotropic and distance-dependent.
In the literature, most force-based pedestrian models include anisotropic mechanisms. 
These mechanisms enable to take vision cone effects \cite{Helbing1995,chraibi2010generalized} 
or preferred crossing direction \cite{Totzeck2020} into account. 
The repulsion terms are weighted by factors depending on the bearing angle $\theta_{ij}$ or other speed-based factors breaking the symmetry. 
For instance, in the 
generalized centrifugal force model \cite{chraibi2010generalized}, 
the weight is
\begin{equation}
    \omega_1(\theta_{ij})=\left\{\begin{array}{ll}
    \cos(\theta_{ij})&\text{if }|\theta_{ij}|<\pi/2\\[1mm]
    0&\text{otherwise.}
    \end{array}\right.
\end{equation}
Anisotropic effects also include adaptation of the velocity to the local environment in direction of motion.
Such a phenomenological relation between speed and local density is related to the fundamental diagram in traffic engineering.   
The fundamental diagram is one of the main empirical characteristics of pedestrian dynamics \cite{seyfried2005fundamental,Boltes2018}. 
Altogether, the symmetry of isotropic repulsive interactions is broken by anisotropic effects which we indicate by $\times$ in Fig.~\ref{figPHPMb}.

In contrast, the port-Hamiltonian formulation requires the interaction to be odd, hence anisotropic effects need to be incorporated differently. For example, they can result from state-dependent input terms.
An extended microscopic model reads in this case
\begin{equation}\label{modmicro2}
   \begin{cases}
         ~\dot Q_{ij}(t)=p_i(t)-p_j(t),\qquad\qquad\qquad\qquad Q_{ij}(0)=Q_{ij}^0,\quad p_i(0)=p_i^0,\\[1mm]
         ~\dot p_i(t)= \lambda\big(u(t)V(Q_i(t),p_i(t))-p_i(t)\big)-\sum_{j\not=i}\nabla U\big(Q_{ij}(t)\big),
    \end{cases}
\end{equation}
with $V$ a nonlinear optimal velocity function (fundamental diagram) depending on position differences and current velocity. 
It may also be a function of the velocities of the neighbors to model group effects.
The corresponding port-Hamiltonian formulation is 
\begin{equation}\label{PHS2}
\begin{cases}
    ~\dot{z}(t)=(J-R)\nabla H(z(t))+B(z(t))u(t),\qquad\qquad z(0)=z_0\\[1.5mm]
    ~y(z(t))=B^\text{\normalfont T}(z(t))\nabla H(z(t)),
\end{cases}
\end{equation}
with 
\begin{equation}
B(z(t))=\left[\begin{matrix}0 \\\lambda\,\text{\normalfont diag}(V(Q_1(t),p_1(t)),\ldots,V(Q_N(t),p_N(t)))\end{matrix}\right]
\end{equation}
a dynamical input matrix. 
The PHS still admits an input-state-output structure with no interaction port.
Yet, nonlinear effects come not only from the interaction potentials but also from the input matrix $B$ operating as a dynamical feedback \cite{ortega2001putting}.
The velocity-dependence of $V$ yields an anisotropy and potentially enables to describe fundamental diagrams and other related features such as stop-and-go waves \cite{Boltes2018}. 
The challenge (as illustrated in Fig.~\ref{figPHPMb} with the question mark) consists of formulating realistic anisotropic mechanisms through the optimal velocity function $V$. 
For instance, the optimal velocity may solely depend on the distance in front \cite{tordeux2016collision,xu2019generalized} or on a preferred crossing side \cite{Totzeck2020}. 
Other velocity functions may depend on the velocities of the neighbors to model swarming behavior of pedestrian groups. 

\subsubsection*{Acknowledgements} 
The authors gratefully acknowledge Prof.\ Andreas Frommer, Prof.\ Michael G\"unther, and Dr.\ Karsten Kahl for meaningful suggestions and support in the formulation and analysis of the numerical schemes. The authors also gratefully acknowledge Prof.\ Andreas Schadschneider for interesting discussions and relevant remarks regarding the use of the Hamiltonian as order parameter.

\subsubsection*{Data and results accessibility} 
An online simulation platform of the port-Hamiltonian pedestrian model on a torus is available at
\bc\medskip\href{https://www.vzu.uni-wuppertal.de/fileadmin/site/vzu/Port-Hamiltonian_pedestrian_model.html?speed=0.5}{\texttt{https://www.vzu.uni-wuppertal.de/fileadmin/site/vzu/ Port-Hamiltonian\_pedestrian\_model.html}}.\medskip\ec
The readers can initiate a simulation and setting $\lambda=0$, $u=0$, or $A=0$ to observe dynamics similar to those presented in Figs.~\ref{fig:Trajl0}, \ref{fig:Traju0} and \ref{fig:TrajA0}, respectively.  
The same parameters in combination with scenario \textit{counter-flow} lead to lane formation as in Fig.~\ref{fig:TrajLF} or to \textit{crossing-flow} for strip formation as in Fig.~\ref{fig:TrajSF}. 
The gridlocks shown in Figs.~\ref{fig:TrajLFb} and \ref{fig:TrajSFb} are obtained for small $\lambda$.



\end{document}